\documentclass[10pt,conference]{IEEEtran}

\usepackage{amsmath,amssymb,amsfonts}
\usepackage{algorithmic}
\usepackage{graphicx}
\usepackage{textcomp}
\usepackage{xcolor}
\usepackage{colortbl}
\usepackage{xspace}
\def\BibTeX{{\rm B\kern-.05em{\sc i\kern-.025em b}\kern-.08em
    T\kern-.1667em\lower.7ex\hbox{E}\kern-.125emX}}

\usepackage[numbers,sort]{natbib}
\usepackage{fancyhdr}
\usepackage{hyperref}
\usepackage[nameinlink, capitalize, noabbrev]{cleveref}
\usepackage{enumitem}
\usepackage{multirow}
\usepackage{tikz}
\usepackage{array}
\usepackage{booktabs}
\newcolumntype{Y}{>{\raggedleft\arraybackslash}p{9.6mm}}
\usepackage{tablefootnote}
\usepackage{adjustbox}
\usepackage{float}
\usepackage{amsthm}
\usepackage{threeparttable}
\usetikzlibrary{shapes.geometric}

\usepackage[most]{tcolorbox}
\usepackage{soul}

\newcommand*\circled[1]{\tikz[baseline=(char.base)]{
            \node[shape=circle,draw,inner sep=0.2pt] (char) {#1};}}
            
\definecolor{obsblue}{RGB}{31,78,121}   %
\definecolor{obsback}{RGB}{240,244,249} %
\newtcolorbox{observation}{
    enhanced,
    colback=obsback,
    frame hidden,
    boxrule=0pt,
    borderline west={2.2pt}{0pt}{obsblue},
    arc=0.6mm,
    left=6pt,
    right=5pt,
    top=2pt,
    bottom=2pt,
    before skip=4pt,
    after skip=4pt,
    fontupper=\normalfont
}

\definecolor{darkviolet}{HTML}{9400D3}
\definecolor{darkgreen}{rgb}{0,0.65,0}
\definecolor{cacheCircleBlue}{RGB}{8,131,202}

\definecolor{skew}{RGB}{102,181,224}
\definecolor{mirage}{HTML}{B5C3C3}
\definecolor{ceaser}{HTML}{ECE69D}
\definecolor{idf}{HTML}{A4CF5C}

\DeclareRobustCommand{\circnum}[2]{%
  \tikz[baseline=(char.base)]{
    \node[
      shape=circle,
      draw=black,
      fill=#2,
      text=black,
      line width=0.35pt,
      inner sep=0.6pt,
      minimum size=1.1em,
      font=\scriptsize
    ] (char) {#1};
  }%
}

\newcommand{\fopt}{\textsuperscript{\textit{F}}}

\newcommand{\cache}[1]{\textsc{#1}}
\newcommand{\ceaser}{\cache{Ceaser}\xspace}
\newcommand{\ceasers}{\cache{Ceaser\mbox{-}S}\xspace}
\newcommand{\ceasersf}{\cache{Ceaser\mbox{-}S}\fopt\xspace}
\newcommand{\scattercache}{\cache{Scat\-ter\-Cache}\xspace}
\newcommand{\sasscache}{\cache{Sass\-Cache}\xspace}
\newcommand{\sasscachef}{\cache{Sass\-Cache}\fopt\xspace}
\newcommand{\mirage}{\cache{Mirage}\xspace}
\newcommand{\miragef}{\cache{Mirage}\fopt\xspace}
\newcommand{\std}{\cache{STD}\xspace}

\newcommand{\stda}{\cache{STD\mbox{-}3C}\xspace}

\newcommand{\riscv}{RISC\mbox{-}V\xspace}

\newcommand{\bench}[1]{\texttt{#1}}

\hypersetup{
        hyperfootnotes=true, %
        colorlinks,
        linkcolor={red!50!black},
        citecolor={darkgreen!70!black},
        urlcolor={blue!80!black}
}

\title{Randomized Caches in Hardware: \\ Challenges and Trade-offs}

\newcommand{\hpcaauthors}{Xinrui Wang\IEEEauthorrefmark{1}, Pascal Kosak\IEEEauthorrefmark{1}, Christof Paar\IEEEauthorrefmark{2}, Daniel Genkin\IEEEauthorrefmark{3}, Yuval Yarom\IEEEauthorrefmark{1}, Anirban Chakraborty\IEEEauthorrefmark{2}}

\newcommand{\hpcaaffiliation}{\\
  \IEEEauthorrefmark{1} Ruhr University Bochum \quad
  \IEEEauthorrefmark{2} Max Planck Institute for Security and Privacy  \quad
  \IEEEauthorrefmark{3} Georgia Tech
}
\newcommand{\hpcaemail}{\relax}

\author{
    \IEEEauthorblockN{\hpcaauthors{}}
      \IEEEauthorblockA{
        \hpcaaffiliation{} \\
        \hpcaemail{}
      }
}

\fancypagestyle{camerareadyfirstpage}{%
  \fancyhead{}
  
  \fancyfoot[C]{}
}
\begin{document}
\maketitle

\thispagestyle{camerareadyfirstpage}
\pagestyle{empty}

\bstctlcite{IEEEexample:BSTcontrol}

\begin{abstract}

Multiple recent works propose \emph{randomized cache} designs, that protect against side-channel attacks by obfuscating the mapping of memory addresses to cache locations.
These designs typically specify the architectural behavior of the scheme, leaving the practical challenges of a concrete hardware implementation largely unexplored.
Moreover, in the absence of hardware implementations, designers have to resort to architectural simulators for evaluating the design, introducing a further level of abstraction.
To bridge the gap between the abstract designs and their concrete realization, we implement four prominent randomized cache designs---\ceasers, \scattercache, \sasscache and \mirage---and integrate them as the last-level cache (LLC) of a multi-core out-of-order \riscv system-on-chip (SoC).
We first identify multiple common challenges to implementation, as well as several design-specific ones, and  develop practical solutions to overcome them.
We then evaluate the resulting SoC variants through complete FPGA implementations on which we run the SPEC CPU 2017 rate and GAPBS workloads to completion, presenting, to our knowledge, the first comparative evaluation of these designs integrated into a common full-system hardware platform.
Our findings indicate that pipelining the index derivation function (IDF), as suggested by past works, is not sufficient.
In three of the four variants, the frequency bottleneck shifts to other aspects of the design, resulting in a frequency drop of up to 35\% compared to the baseline set-associative cache.
Redesigning the critical paths, while keeping each design's mappings, rekeying, and victim-selection policies unchanged, brings the three affected variants to not more than 10\% below the baseline frequency, at the cost of added latency or reduced concurrency. 
We further identify that increased LLC access latency, rather than additional misses, dominates the number of instructions executed per cycle (IPC), reducing it by up to 5\% for SPEC and 11\% for GAPBS.
The resource overhead varies significantly based on the specific resource, the design, and the implementation technology. For FPGA, we see an increase of 10--35\% in the number of lookup tables (LUTs), whereas for ASIC, we see an area increase of 1--14\%.

\end{abstract}


\section{Introduction}

Modern processors rely on caches to hide the long latency of memory accesses.
However, the timing difference between a cache hit and a miss has become a fertile ground for side-channel attacks~\cite{kocher1996timing,page2002theoretical,bernstein2005cache,osvik2006cache,yarom2014flush}.
This is especially true for last-level caches (LLCs), which are shared across cores and can be exploited to leak information across security domains~\cite{liu2015last, maurice2017hello, IrazoquiES15, ZhaoMFT24}.
Contention-based attacks exploit this timing difference by filling selected cache sets with attacker-controlled lines and then measuring whether accesses to those lines have become slower, which reveals that the victim evicted them.
Research over the last two decades has shown that such attacks can leak cryptographic keys~\cite{osvik2006cache, liu2015last, GenkinPTY18}, keystrokes and user behavior~\cite{ristenpart2009hey, OrenKSK15, shusterman2019robust, OConnellSMGOSY24}, and other sensitive information~\cite{ShahverdiSD21, YanFT20, LeeRKLH26, HundWH13}.
Central to these attacks is the construction of minimal eviction sets: collections of addresses that map to the same cache set and can collectively displace a target line~\cite{liu2015last, ZhaoMFT24, qureshi2019new, VilaKM19, KessousG24, MorganHOSCGMMRY25, PurnalBPV23, PurnalTV21, ThomaG22}. %
Constructing eviction sets is practical in conventional caches because 
the mapping of addresses to cache sets predictably partitions memory addresses based on address bits.

\begin{figure}[!t]
    \centering
    \includegraphics[scale=0.72]{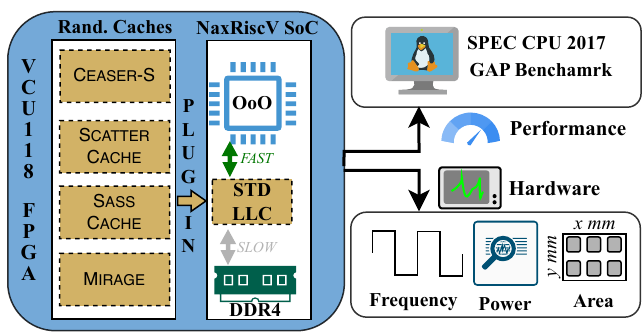}
    \caption{Architectural overview of randomized cache implementation.}
    \label{fig:weDone}
    \vspace{-2mm}
\end{figure}

Randomized caches counter contention-based attacks by breaking this fixed mapping.
Early designs randomized the address-to-cache mapping through lookup tables~\cite{wang2007new,wang2008novel} or randomized the fill policy~\cite{liu2014random}, but these approaches target small L1 caches and scale poorly to large LLCs.
Instead, more recent designs~\cite{qureshi2018ceaser, qureshi2019new, werner2019scattercache, giner2023scatter, saileshwar2021mirage,   tan2020phantomcache, unterluggauer2022chameleon, song2024seqass} compute cache indices with keyed cryptographic functions, called index-derivation functions (IDFs), 
breaking the direct mapping of addresses to cache sets. 
Past work has extensively evaluated the security of randomized caches~\cite{chakraborty2025systematic, BhatlaBSP25, BourgeatDYTEY20, GenkinKLTUY23, PurnalGGV21, PetersSTGY25, SongLXLWL21}, highlighting weaknesses and proposing countermeasures.

Although these proposals describe the high-level design, they stop short of a complete hardware implementation. 
Existing work focuses primarily on a few isolated implementation concerns. 
For example, most designs recognize that the IDF must be pipelined to meet timing constraints and estimate the resulting increase in cache access latency. 
However, integration challenges beyond the IDF, such as critical-path analysis, control pipelining, hazard management and miss-status holding register (MSHR) locking, have received less attention.

In the absence of concrete implementations, performance evaluations primarily rely on architectural simulators~\cite{binkert2011gem5, GoberCWGJTPK22, MaWS26, Spike}.
Although these simulators are invaluable for early stage exploration, they tend to abstract away many implementation issues, resulting in inaccuracies~\cite{akram2019validation,huppert2021memory,walker2018hardware,gutierrez2014sources}. 
Moreover, the speed of simulation with these tools severely limits the amount of code that can be simulated.
Consequently, evaluations only span short sequences of instructions, typically in the low billions or fewer~\cite{saileshwar2021mirage, chakraborty2025systematic, BhatlaBSP25}.
Hardware cost evaluation faces similar limitations. Without synthesized implementations, area, timing, and energy are typically estimated using analytical models, such as CACTI~\cite{muralimanohar2007optimizing}, rather than being derived from an implemented design.

Two works implement variants of \ceaser~\cite{song2024randomizing} and of \scattercache~\cite{curlin2026open}.
Both target the in-order RocketChip core~\cite{rocketchip}, leaving open how these designs translate to high-performance cores with speculative, out-of-order execution. 
Moreover, since each work implements only one design, a direct comparison of the implementation trade-offs across designs remains unavailable.
This raises the following questions:
\circled{1}~\textit{What are the challenges and trade-offs for randomized cache implementation?}
\circled{2}~\textit{What do they cost in clock frequency, performance, and area once fully integrated?}

In this work, we answer these questions through an end-to-end study, illustrated in \cref{fig:weDone}.
In the first step, we select four prominent randomized cache designs, \ceasers{}~\cite{qureshi2019new}, \scattercache{}~\cite{werner2019scattercache}, \sasscache{}~\cite{giner2023scatter}, and \mirage{}~\cite{saileshwar2021mirage}, that highlight different aspects of the design space.
We identify several challenges to implementing these---some stem from the skew cache design~\cite{Seznec93} that most modern randomized caches build on, whereas others are introduced by the relocation and indirection aspects of some specific designs.
We propose solutions for these challenges and develop RTL implementations of all four cache designs.

We integrate the different cache design as the LLC of a multi-core out-of-order NaxRiscv~\cite{spinalhdl2026naxriscv} RISC-V SoC,   and synthesize the resulting systems on FPGA.
The synthesized SoCs allow us to characterize the physical properties of the implementations, including maximum operating frequencies, resource utilization, and power consumption.
Moreover, by booting Linux on the synthesized SoCs, we execute complete benchmark suites, such as SPEC CPU2017~\cite{spec2017spec} and GAPBS~\cite{beamer2015gap}.
The execution speed of the FPGA allows these benchmarks to run to completion---averaging 290 billion instructions per benchmark, with some exceeding one trillion instructions---thereby measuring the performance impact of cache randomization over complete application executions rather than short simulated instruction windows.

We observe that, except for \sasscache, whose dynamic partitioning reduces effective cache capacity, randomized-cache designs incur similar instructions-per-cycle (IPC) losses.
The loss correlates strongly with the LLC access rate and reaches up to $10\%$ for memory-intensive workloads, while miss-related effects remain secondary.
Adding three cycles of latency---matching the IDF latency of \scattercache and \mirage---to a standard set-associative cache reproduces most of IPC loss.
Together, these observations indicate that the latency introduced by the IDF, rather than increased cache misses, is the dominant factor for IPC degradation.

We further observe that the IDF is not the only source of signal propagation delays.
Even with a pipelined IDF, all designs exhibit a frequency drop ranging from $1\%$ for \scattercache to $35\%$ for \mirage, an effect that has been so far overlooked.
Critical-path analysis reveals that asynchronous operations, such as relocation in \ceasers and indirection in \mirage are the primary source for large frequency drops.
We redesign \sasscache, \ceasers, and \mirage to mitigate these critical paths, recovering much of the lost frequency at the expense of additional LLC latency or cache locking. Overall, after accounting for both IPC degradation and frequency reduction, randomized caches reduce the maximum execution throughput by up to $35\%$.

\smallskip
\noindent In summary,  we make the following contributions:
\begin{itemize}[nosep,left=0pt]
\item We provide the first full implementations of advanced SoCs supporting randomized caches (\cref{sec:challenges,sec:implementation}).
\item We identify and analyze the causes of frequency drops in the designs and show how to overcome them (\cref{sec:redesign}).
\item We perform end-to-end comparison of the performance and the resource usage of the designs (\cref{sec:evaluation}).
\end{itemize}

\section{Background}
\label{sec:background}

\subsection{Cache and Cache Attacks}

Caches in modern processors are typically organized as set-associative structures, where each memory address is mapped to a (fixed) subset of cache lines, called a set, and each set contains multiple ways (or cachelines) that can hold different data. 
Addresses that are mapped to the same set are said to be \emph{congruent} and can evict each other from the cache.
When loading a new line into the cache, the replacement policy determines which of the existing lines in the set to evict. 
Each cacheline is associated with a tag and additional metadata that uniquely identifies the memory address it corresponds to, as well as its coherence state.

Modern CPUs employ a multi-level cache hierarchy: the L1 (and often the L2) caches are private to each core and are kept consistent by a coherence protocol, whereas the last-level cache (LLC) is shared among cores and is in many cases inclusive of the private levels, i.e., any line present in a private cache is also present in the LLC.
Large LLCs are additionally split into multiple slices~\cite{maurice2015reverse,irazoqui2015systematic}, each associated with a core or group of cores.
As the LLC is shared across cores and thus across security domains, it is a prime target for contention-based side-channel attacks such as Prime+Probe~\cite{osvik2006cache}.
In these attacks, the attacker fills a monitored cache set with an \emph{eviction set}---a collection of congruent addresses occupying all of its ways---and later re-accesses these lines, where increased access latency reveals an intervening victim access to the same set.
Since conventional caches derive the set index directly from address bits, such eviction sets are straightforward to construct~\cite{liu2015last, qureshi2019new, VilaKM19, PurnalTV21, KessousG24, MorganHOSCGMMRY25, PurnalBPV23, ThomaG22}.

\subsection{Secure Cache Designs}
\begin{figure}[t]
    \centering
    \includegraphics[scale=0.9]{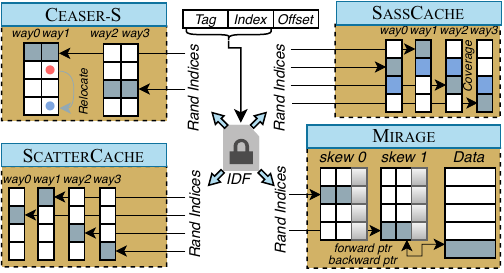}
    \caption{The overview of the randomized caches.}
    \label{fig:cache-class}
\end{figure}

Secure caches make contention attacks statistically hard by breaking the deterministic mapping between addresses and cache sets. 
The most common approach is to use a keyed cryptographic function, called an index derivation function (IDF), to derive the cache set index from the address.\footnote{Earlier designs based on permutation tables~\cite{wang2007new,wang2008novel,trilla2018cache} target small caches and scale poorly to large LLCs, while partitioned caches without randomization~\cite{domnitser2012non,sanchez2011vantage,beckmann2013jigsaw,schwedock2020jumanji} sacrifice flexibility and performance. We do not consider them in this paper.} 
Contemporary randomized caches~\cite{qureshi2019new, werner2019scattercache, tan2020phantomcache, giner2023scatter, saileshwar2021mirage, song2024seqass, BhatlaNQP26, ThomaNFLMPDG23, BhatlaNP24, CzernyLGCG26} employ set- or skewed-associative design with the IDF providing the required randomization. \Cref{fig:cache-class} summarizes the representative designs considered in this work.

\noindent \textbf{\ceasers~\cite{qureshi2019new}} partitions the cache into multiple (typically two) independently indexed skews, each indexed by a separate IDF.\footnote{We consider \ceaser{} as a special case of \ceasers{} with a single skew.} 
A new line is inserted into a randomly selected skew, and eviction is restricted to that same skew. 
To retire learned mappings, \ceasers{} periodically rekeys the IDFs and gradually relocates resident lines using an access counter to amortize the transition between epochs.

\noindent \textbf{\scattercache~\cite{werner2019scattercache}} uses a single IDF shared across all ways. 
Different bit masks of the IDF output are used to derive an independent cache-set index for each way, giving every address one candidate location. 
A new line is inserted into a randomly selected way, and eviction is restricted to that way.

\noindent \textbf{\sasscache~\cite{giner2023scatter}} extends \scattercache{} by introducing a second level of randomization. 
Each way is further partitioned into multiple independently indexed sub-ways, giving each address one candidate location in every sub-way. 
A new line is inserted into a randomly selected sub-way, and eviction is restricted to that sub-way. 
Because each security domain maps to only a subset of the cache locations, \sasscache{} trades effective cache capacity for stronger isolation.

\noindent \textbf{\mirage~\cite{saileshwar2021mirage}} separates the tag store from the data store to approximate fully associative replacement. 
The tag store remains set associative with additional invalid tag entries to avoid set conflicts, while cache misses perform \emph{global eviction} by randomly selecting a data entry, invalidating its corresponding tag, and installing the incoming line. 
Mirage further employs two-skew indexing with load balancing and randomized address mapping via IDF for added security.

\subsection{Hardware Cache Structure}
\label{sec:background-cache-hardware}

\begin{figure}[t]
    \centering
    \includegraphics[scale=0.95]{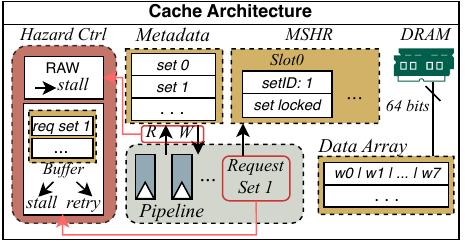}
    \caption{General overview of a hardware last-level cache architecture.}
    \label{fig:cache-background}
\end{figure}

Mature open-source RISC-V processors, including NaxRiscv~\cite{spinalhdl2026naxriscv}, XiangShan~\cite{xiangshan}, and BOOM~\cite{boom}, implement LLCs that share the general architecture shown in \cref{fig:cache-background}; we use NaxRiscv as the running example throughout this paper.
We next describe three properties of this architecture---non-blocking operation, set-granularity management, and decoupled data and metadata storage---that cache randomization directly disrupts.

\subsubsection{Non-Blocking Operation}
Modern LLCs are non-blocking: 
The latency of a memory access in the case of a cache miss can often take more than 100 cycles.
To prevent idling for such a long period, the LLC combines a pipelined access path with miss-status holding registers (MSHRs)~\cite{kroft1981lockup}.
As \cref{fig:cache-background} shows,
 requests traverse a multi-stage pipeline in which operations, such as reading and updating the metadata, are performed in different stages, so consecutive requests are processed back to back.
When a request misses, an MSHR entry records the miss and its target set while the line is fetched from memory; once the response returns, the data is forwarded to the requester and the entry is released.
MSHRs thus allow the cache to keep serving hits and to overlap multiple outstanding misses.

This parallelism, however, introduces hazards, which the cache resolves with a hazard controller.
Two hazards are typical.
A metadata read-after-write (RAW) hazard arises when one pipeline stage reads metadata that another stage is still updating, and hence may observe stale state; the hazard controller stalls the read until the update completes.
In contrast, MSHR conflicts arise when a request targets a cache set with an outstanding miss. 
The open-source LLCs we consider update the metadata eagerly when the MSHR entry is allocated, while the corresponding data arrives from memory later. 
The set therefore remains temporarily inconsistent until the refill completes.
The hazard controller therefore locks the sets recorded in the MSHRs (the \emph{MSHR lock}), buffering conflicting requests or stalling the pipeline, depending on the exact scenario.

\subsubsection{Set-Granularity}
Conventional set-associative caches manage most operations at the granularity of a cache set.
The tags and associated metadata of all of the ways of a set are typically stored in a single wide word of the metadata SRAM, so one access fetches all of them and the hit check compares them in parallel.
Replacement state is likewise maintained per set, for instance by a pseudo-least-recently-used (PLRU) policy, and only cache lines within the same set compete for occupancy.
Set granularity also keeps hazard handling simple. Congruent requests share the same set index, so the hazard controller and the MSHR lock detect conflicts by comparing set indices alone, minimizing detection logic while preserving access parallelism.

\subsubsection{Decoupled Data and Metadata Arrays}
Finally, data and metadata are stored in physically separate SRAM arrays.
As shown in \cref{fig:cache-background}, the memory interface is typically narrower than a cache line ($64$ bits versus $64$-byte lines in NaxRiscv), so a refill arrives over multiple (at least $8$) cycles and occupies the data array for multiple cycles.
Many frequent operations, such as coherence-state and directory updates, access only the metadata.
If data and metadata shared one SRAM structure, these lightweight operations would serialize behind long-running refills.
Separating the arrays lets metadata operations proceed independently of memory-side refill traffic, increasing parallelism between control-path and data-path activity.

\section{Hardware Implementation Challenges}
\label{sec:challenges}

\begin{figure}[t]
    \centering
    \includegraphics[scale=0.9]{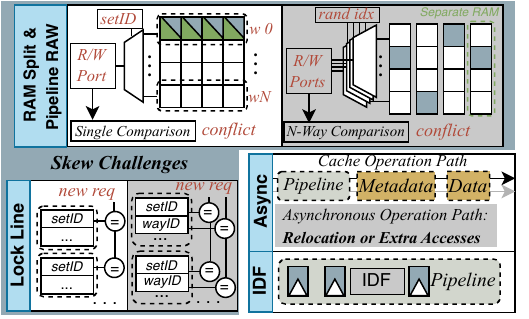}
    \caption{Randomization techniques and their impact on the baseline LLC.}
    \label{fig:challenges}
\end{figure}

We study four representative randomized cache designs: \scattercache, \ceasers, \sasscache, and \mirage.
Although proposed as different schemes, we observe that they build on a shared hardware foundation while differing only in a few design-specific mechanisms. 
This allows us to separate the implementation challenges common to randomized caches from those specific to individual designs.

All four designs share two fundamental mechanisms. First, cache indices are derived using a keyed \textit{index-derivation function (IDF)} rather than directly from the address. 
Second, the cache is organized as \textit{independently-indexed skews} instead of using a single shared index. 
Consequently, every demand access must first compute the randomized indices before performing metadata lookup, hazard detection, replacement, and cache management on independently indexed structures. 
Supporting {IDF computation} and {skewed indexing} is therefore fundamental to contemporary randomized caches.

\scattercache basically implements these two mechanisms, providing the minimal architecture for a randomized cache.
\ceasers{} adds {relocation}, periodically rekeying its IDFs, relocating resident lines to retire learned mappings. 
\sasscache{} extends the lookup hierarchy with independently indexed sub-ways while dynamically partitioning the cache to strengthen occupancy isolation. 
Finally, \mirage{}  introduces {indirect eviction}, selecting a random data-store entry on a cacheline install and invalidating its owning tag through a reverse pointer.
Collectively, these schemes cover a representative range of the mechanisms used in state-of-the-art randomized caches and explore the performance impact of various features.

From a hardware perspective, these mechanisms naturally divide into two classes. IDF computation and skewed metadata lookup execute on every demand access and therefore form the \emph{synchronous demand path}. 
In contrast, relocation and indirect eviction execute independently of demand requests, overlapping with and contending for the same  resources as normal cache traffic.
We refer to these as \emph{asynchronous paths}. \Cref{fig:challenges} summarizes these mechanisms and the implementation challenges they introduce.
We now describe these challenges.

\subsection{Handling Cache Skew}
\label{sec:cache-skew}

Skewing affects the baseline LLC's metadata access and conflict management in three ways.
First, the baseline reads all ways of a set from a single wide metadata-RAM word, whereas a skewed cache accesses an independent row in each skew. 
Retaining the original single-read-port organization would therefore serialize metadata lookup into one access per independent index---two for \ceasers{} and \mirage{}, and one per way for \scattercache{} and \sasscache{}---increasing LLC latency.
Second, a single set-index comparison no longer detects all metadata read-after-write (RAW) hazards, since requests conflict whenever they access the same row of the same metadata bank, even if their original set indices differ.
Third, MSHR locking can no longer be performed at set granularity because requests to different metadata banks may proceed independently despite sharing the same row index.

\noindent\textbf{Our Solution.}
We partition the wide metadata RAM into independently accessible banks, restoring single-cycle metadata lookup through parallel reads. 
RAW hazards are detected independently for each skew, and each MSHR entry records both the selected way and row index. 
This line-level lock preserves parallel lookup and non-blocking operation without unnecessary set-level serialization.

\subsection{IDF Integration}
\label{sec:idf}
The IDF must generate the randomized index before metadata lookup and therefore lies directly on the LLC access path. 
Because the IDF is implemented as a keyed hash or block cipher, a fully combinational design lengthens the critical path and reduces the maximum operating frequency.
Prior work addresses this by pipelining the IDF, e.g., over three stages for \mirage~\cite{saileshwar2021mirage}.
Pipelining shortens the combinational delay and, once the pipeline is filled, sustains one index per cycle, but introduces additional latency before metadata lookup.

\noindent\textbf{Our Solution.}
Following the original proposals, we implement each IDF as a fully pipelined datapath placed before the metadata RAM.
This removes the IDF from the single-cycle critical path while preserving one-request-per-cycle throughput, at the cost of a design-specific pipeline latency.

\begin{figure*}[t]
    \centering
    \includegraphics[scale=0.8]{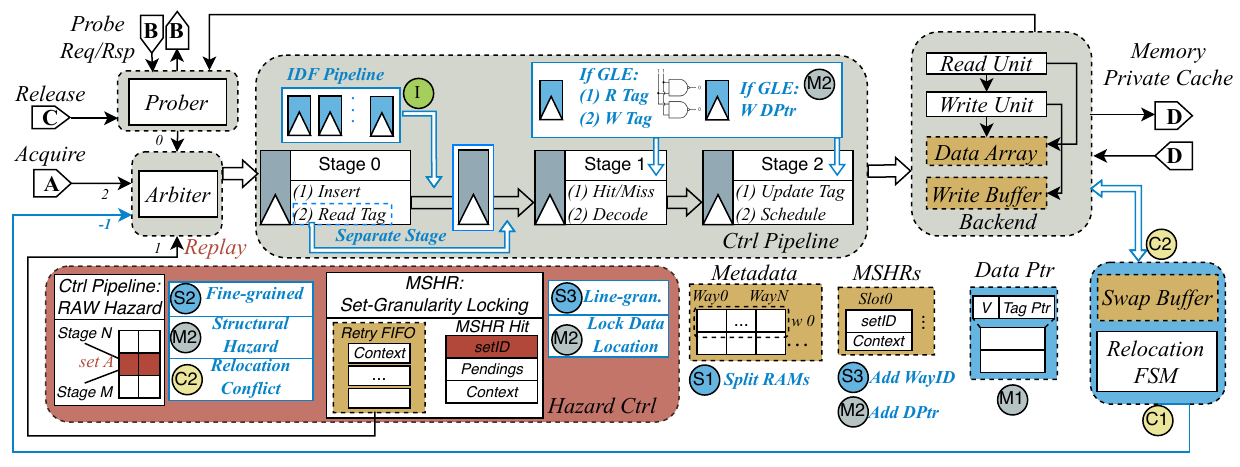}
    \caption{Randomized LLC implementations integrated into the NaxRiscV SoC~\cite{spinalhdl2026naxriscv}. \circnum{S}{skew} denotes design elements shared by all randomized LLCs to solve cache skewing, \circnum{I}{idf} denotes the IDF integration, and \circnum{C}{ceaser} and \circnum{M}{mirage} denote elements specific to \ceasers and \mirage for asynchronous operations, respectively.}
    \label{fig:cache-structure}
\end{figure*}

\subsection{Asynchronous Paths}
\label{sec:async-paths}
Unlike skew and IDF integration, asynchronous paths are not part of the demand lookup. 
Instead, they execute as multi-cycle operations alongside regular requests, and compete for cache resources.
Although \ceasers{} relocation and \mirage{} global eviction serve different purposes, both introduce conflicts absent from the baseline LLC. 
In \ceasers{}, a demand request may target a line being relocated; the request must not observe this transient state and therefore stalls until relocation completes.
In \mirage{}, global eviction generates additional dependent accesses to the metadata and pointer arrays. These accesses may contend with the demand pipeline for array ports or reach an in-flight data entry through a different metadata entry, creating both structural and data-location hazards.

\noindent\textbf{Our Solution.}
We route these asynchronous paths through the existing control pipeline and overlap them with regular requests instead of serializing all demand traffic.
Hazard detection tracks only the affected locations and array accesses so that only conflicting operations are stalled, replayed, or arbitrated, while independent requests proceed.
This selective synchronization preserves correctness with minimal performance impact (cf.\ \cref{sec:ceaser-impl,sec:mirage-impl}).

\section{End-to-end Implementation}
\label{sec:implementation}
Having identified the platform-independent implementation challenges in \cref{sec:challenges}, we now realize the four caches as complete end-to-end hardware implementations.
We implement all four designs in NaxRiscv~\cite{spinalhdl2026naxriscv}, an open-source RISC-V SoC comprising an out-of-order core and a coherent L2 cache that serves as the LLC.
NaxRiscv is optimized for both area efficiency and high clock frequency, making it a demanding platform for evaluating the timing and hardware implications of cache randomization.
Wherever possible, we preserve the original organization while adapting the implementation to the NaxRiscv cache hierarchy.
Our objective is to preserve the throughput of the baseline LLC and its non-blocking operation while minimizing additional latency.
We first present the mechanisms common to all four implementations, followed by the design-specific mechanisms required for \ceasers{} and \mirage{}.

\subsection{Common Hardware Changes}
\label{sec:common-impl}

All  implementations share four hardware changes, highlighted in \cref{fig:cache-structure}. %
Changes \circnum{S1}{skew}, \circnum{S2}{skew}, and \circnum{S3}{skew} address the skew-related challenges of \cref{sec:cache-skew}, whereas \circnum{I}{idf} integrates the pipelined IDF of \cref{sec:idf}.
Their realization depends on each design's IDF and the baseline LLC organization.

We first partition the metadata RAM to match each design's indexing function (\circnum{S1}{skew}).
Because \scattercache{} and \sasscache{} independently index every way, their metadata is organized as $N$ independently addressed way RAMs, allowing all candidate tags to be accessed in parallel.
In contrast, \ceasers{} and \mirage{} employ two skews.
We therefore partition the metadata RAM into two independently addressed banks.
The data array remains unchanged because randomization only affects metadata lookup.

Next, we refine pipeline hazard detection from the set level to the granularity of an independent metadata index (\circnum{S2}{skew}): per way for \scattercache{} and \sasscache{}, and per skew for \ceasers{} and \mirage{}.
Finally, we replace the baseline set-level MSHR lock with a line-level lock (\circnum{S3}{skew}), extending each MSHR entry to record the complete metadata location of an in-flight cache line.
For \ceasers{}, \scattercache{}, and \sasscache{}, this refinement eliminates unnecessary serialization between independent accesses.
For \mirage{}, it additionally ensures correctness when multiple metadata entries  reference the same physical cache line.

We integrate the IDF as a fully unrolled, pipelined datapath placed immediately before the metadata RAM (\circnum{I}{idf}).
Following the original proposals, we implement a two-stage Feistel network for \ceasers{}, a three-stage QARMA~\cite{avanzi2017qarma} pipeline for \scattercache{}, the original three-stage \sasscache{} permutation, and a three-stage PRINCE~\cite{borghoff2012prince} pipeline for \mirage{}.
Since metadata lookup depends on the randomized index, it cannot begin until the IDF completes.
Consequently, an $N$-stage IDF inserts $N$ additional stages before the baseline metadata lookup stage, shifting metadata lookup to stage $N+1$.
Nonetheless, once the pipeline is filled, all four implementations can sustain one randomized index per cycle, preserving the baseline request throughput.

\subsection{\ceasers-specific Implementation}
\label{sec:ceaser-impl}

\ceasers{} preserves its security guarantee by periodically rekeying the cache and incrementally relocating one cache set at a time instead of flushing the entire cache.
Realizing this mechanism in hardware introduces two implementation challenges: tracking each relocation until completion and maintaining correctness in the presence of concurrent requests.

\subsubsection{Status Tracking}

Relocation is a multi-cycle operation whose latency depends on the destination state.
If the destination is invalid, relocation simply transfers the line's data and metadata.
If the destination is valid, however, it must first be evicted by issuing coherence probes to the private caches, obtaining the most up-to-date copy, completing the eviction, and only then moving the relocated line's data and metadata.
The hardware must therefore orchestrate this variable-latency operation and determine when relocation has completed before initiating the next one.

We assign this responsibility to a dedicated relocation finite-state machine (FSM) (\circnum{C1}{ceaser}), which coordinates the relocation sequence and tracks the state of the line being relocated.
Rather than introducing a dedicated control pipeline, the FSM reuses the existing cache-control pipeline (\cref{fig:cache-structure}), minimizing both hardware and control overhead.
It injects relocation requests through the pipeline arbiter, while the MSHRs track each relocation until completion.
Unlike prior work that relocates lines sequentially for \ceaser~\cite{song2024randomizing}, our FSM exploits the available cache concurrency by scheduling and processing as many relocation requests in parallel as resources permit.

\subsubsection{Concurrent Request Handling}

To preserve performance, \ceasers{} performs non-blocking relocations, allowing the cache to continue serving regular requests while a relocation is in progress.
This introduces potential conflicts between an in-flight relocation and demand requests.
A regular request that targets a line being relocated must not observe stale data or an intermediate relocation state, and therefore cannot proceed until relocation completes.
Coherence requests on the {\textbf{C}} channel are more challenging.\footnote{NaxRiscv uses the TileLink~\cite{cook2016productive} coherence protocol. Channels \textbf{A} and \textbf{D} carry requests, responses, and data, while \textbf{B} and \textbf{C} handle coherence probes, acknowledgments, and writebacks.}
Private caches use this channel to evict and write back cache lines, it provides no retry mechanism, and must be prioritized to preserve coherence correctness.
Consequently, the cache cannot stall, reject, or replay conflicting \textbf{C}-channel requests.

We extend the hazard controller (\circnum{C2}{ceaser}) to coordinate both relocation timing and relocation-induced conflicts.
Once the relocation counter reaches its threshold, the controller searches the cache pipeline, request queues and MSHRs for outstanding accesses that conflict with the target set.
Relocation begins only after all conflicting accesses have completed, allowing the controller to identify a safe execution window without draining the cache.
Conflicts arising during relocation are then resolved according to request type.
Regular requests targeting a relocating line are redirected to the retry FIFO and replayed after relocation completes.
In contrast, conflicting \textbf{C}-channel requests are captured by the swap buffer (\cref{fig:cache-structure}, \circnum{C1}{ceaser}), and their data are merged with the relocating line.
Relocation therefore operates on the most recent copy while preserving coherence correctness.

\subsection{\mirage-specific Implementation}
\label{sec:mirage-impl}

\mirage{} decouples metadata placement from data placement to emulate a pseudo-fully-associative cache.
This organization requires additional storage, including extra metadata ways and two indirection structures: \emph{tag-pointer} and \emph{data-pointer} arrays. 
Integrating these structures into the cache pipeline requires additional metadata and pointer-array accesses while introducing new hazards arising from the metadata--data indirection.

\subsubsection{Additional Metadata Accesses}

Global eviction in \mirage performs several dependent accesses to the metadata and the pointer array.
The cache first reads the metadata array.
if it determines that global eviction is necessary, it reads the data-pointer array to identify the victim data entry and then performs a second metadata read to retrieve the victim's tag and coherence state.
Finally, it then evicts the selected line, invalidates its metadata entry, and installs the incoming line, writing the new metadata and updating the data-pointer array to reconstruct the metadata--data mapping.
Overall, global eviction performs two metadata reads, two metadata writes, one data-pointer read, and one data-pointer write.
That is, global eviction introduce a metadata read and a write on top of the accesses that conventional cache performs.
Scheduling these dependent accesses without extending the cache access latency is our primary challenge.

We take two measures to address this challenge.
We first separate the data-pointer array from the data array (\cref{fig:cache-structure}, \circnum{M1}{mirage}), allowing pointer accesses to execute in the control pipeline alongside metadata accesses.
This avoids contention between pointer operations and data-array accesses, 
which would be particularly costly during multi-cycle cacheline refills, which would otherwise delay pointer operations.

Second, because each storage array provides one read and one write port,\footnote{1R1W is a common two-port SRAM organization~\cite{YabuuchiSSITTN16}; for e.g., iCE40 FPGAs provide dedicated RW ports instead of symmetric 2RW access~\cite{lattice2020memory}.}
the additional accesses fit within the existing pipeline provided no pipeline stage issues a second read or write to the same array.
Accordingly, the additional metadata read and write are scheduled alongside Stage~1, while the data-pointer update overlaps with Stage~2 (cf.~\cref{fig:cache-structure}).
Unlike the original \mirage{} implementation~\cite{saileshwar2021mirage}, which budgets an additional cycle of tag-access latency, this organization introduces no additional pipeline stage and leaves the normal cache-access latency unchanged.

\subsubsection{Additional Hazards}

The additional array accesses and the decoupled metadata--data organization introduce two new hazards.
The first is a structural hazard caused by contention for the metadata-array ports.
Different pipeline stages may access the metadata array in the same cycle---for example, Stage~0 and Stage~1 both issuing a read---and because the array provides only one read port and one write port, two simultaneous reads or two simultaneous writes cannot proceed together and must therefore be arbitrated.
The second hazard arises from the indirection between the metadata and data arrays.
During global eviction, one operation may select a data entry that another in-flight operation is already accessing.
Because different metadata entries may reference the same physical data entry, the common set- and line-level hazard checks do not detect the conflict.

We extend the hazard controller (\circnum{M2}{mirage}) to handle both cases.
For structural hazards, the controller tracks metadata-array accesses across the pipeline and stalls one stage whenever two stages simultaneously require the same metadata-array port.
For data-location hazards, each MSHR additionally records and locks the physical data-array location of the in-flight cacheline.
Subsequent requests targeting the same location are stalled or redirected to the retry FIFO until the lock is released.

\section{The Frequency Cost of Randomized Caches}
\label{sec:redesign}

The LLC implementations described in the previous section are functionally complete and preserve the original randomized cache organizations. 
However, like prior evaluations of randomized caches, they implicitly assume that integrating a randomized LLC does not affect the processor's maximum clock frequency. 
We now test this assumption on complete hardware implementations. 
We first describe our experimental platform and timing methodology (\cref{sec:conf-exp}) and characterize the IDF in isolation (\cref{sec:IDF_Iso}), the component most commonly regarded as the primary frequency bottleneck. 
We then measure the maximum clock frequency of the complete SoC variants (\cref{sec:freq-loss}), identify the critical paths responsible for the observed frequency loss (\cref{sec:freq-diagnosis}), and redesign them to recover the lost frequency (\cref{sec:freq-restore}).

\subsection{Experimental Setup}
\label{sec:conf-exp}

We configure the NaxRiscv SoC with a single core, separate $32$~KB L1 instruction and data caches, and a shared $2$~MB, $16$-way LLC, matching a typical Intel LLC slice~\cite{DoweckKLMRRRYY17}. 
The baseline SoC (\std) uses the original NaxRiscv LLC with PLRU replacement policy.
We adopt the default parameters of each randomized cache:  \ceasers and \mirage use two skews, the IDF pipeline depths follow  \cref{tab:idf-pipeline-unrolled}, \mirage provisions six additional metadata ways per skew, and \ceasers uses a 1\% relocation rate. 
All randomized caches use random replacement, which provides comparable performance~\cite{chakraborty2025systematic}. 
An LFSR~\cite{forencich2016verilog} supplies the random values for replacement, \mirage{} global eviction, and \ceasers{} rekeying.

We implement both the isolated components and the complete SoCs in Vivado 2020.2 targeting the AMD Xilinx VCU118 (Virtex UltraScale+)~\cite{amd2023vcu118}. 
All SoC variants differ only in the LLC implementation. 
To determine the maximum achievable clock frequency (\(F_{\max}\)), we follow the iterative methodology recommended in AMD's UltraFast Design Methodology Guide~\cite{amd2026ultrafasta}. 
For each variant, we independently rerun synthesis and full implementation from identical RTL at target frequencies ranging from $150$ to $220$\,MHz in $10$\,MHz increments using the \texttt{PerformanceOptimized} synthesis and \texttt{Explore} implementation directives with post-route physical optimization.
Rather than stopping at the first timing failure, we evaluate every target frequency and estimate  \(F_{\max}=1/(T_{\mathrm{target}}-\mathrm{WNS})\), where \(T_{\mathrm{target}}\) is the target clock period and WNS is the  worst negative slack,\footnote{WNS (worst negative slack) is the minimum timing slack over all timing paths. 
A non-negative WNS indicates timing closure at the target clock period, whereas a negative WNS indicates a timing violation.}
and report the highest estimated \(F_{\max}\) across the sweep.

\begin{table}[t]
	\centering
	\caption{Maximum operating frequency of isolated IDFs and wrappers.}
	\label{tab:idf-pipeline-unrolled}
	\footnotesize
	\setlength{\tabcolsep}{3.5pt}
	\begin{adjustbox}{max width=\columnwidth}
		\begin{tabular}{@{}l c l l rr rr@{}}
			\toprule
			\multirow{2}{*}{LLC Design} & \multirow{2}{*}{Stages} & \multirow{2}{*}{Scope} & \multirow{2}{*}{Details} & \multicolumn{2}{c}{Unrolled} & \multicolumn{2}{c}{Pipeline} \\
			\cmidrule(lr){5-6}\cmidrule(lr){7-8}
			& & & & \shortstack[r]{Critical \\Path (ns)} & \shortstack[r]{$F_{\max}$\\(MHz)} & \shortstack[r]{Critical \\Path(ns)} & \shortstack[r]{$F_{\max}$\\(MHz)} \\
			\midrule
			\multirow{2}{*}{\begin{tabular}{l}\scattercache{}\\QARMA$_5$\end{tabular}}
			& \multirow{2}{*}{3}
			& IDF     & 1 IDF & 6.953  & {143.4} & 3.745 & {265.7}  \\
			& & Wrapper & 3 IDFs + slicing & 7.746  & 128.8          & 4.104 & 242.5          \\
			\midrule
			\multirow{2}{*}{\begin{tabular}{l}\sasscache{}\\Ext. QARMA$_5$\end{tabular}}
			& \multirow{2}{*}{3}
			& IDF     & 1 IDF & 8.212  & {121.5} & 4.472 & {222.7}  \\
			& & Wrapper & 3 IDFs + slicing & 10.706 & 93.2           & 4.706 & 211.6          \\
			\midrule
			\multirow{2}{*}{\begin{tabular}{l}\ceasers{}\\Feistel-Network\end{tabular}}
			& \multirow{2}{*}{2}
			& IDF     & 1 IDF (2 halves) & 3.391 & {293.3} & 2.466 & 402.6          \\
			& & Wrapper & 2 IDFs + mux & 3.487 & 285.3          & 2.402 & {413.2}  \\
			\midrule
			\multirow{2}{*}{\begin{tabular}{l}\mirage{}\\PRINCE\end{tabular}}
			& \multirow{2}{*}{3}
			& IDF     & 1 IDF & 6.674 & {149.4} & 3.533 & {281.5}  \\
			& & Wrapper & 1 IDF + slicing & 6.688 & 149.1          & 3.804 & 261.6          \\
			\bottomrule
		\end{tabular}
	\end{adjustbox}
\end{table}

\subsection{The IDF in Isolation}
\label{sec:IDF_Iso}

Prior work~\cite{saileshwar2021mirage,giner2023scatter,werner2019scattercache} identifies the IDF as the potential frequency bottleneck in complete SoC implementations.  
Before evaluating the full-SoC, we therefore characterize the four default IDFs in isolation by measuring the maximum operating frequency of both the pipelined IDF and its wrapper (\cref{tab:idf-pipeline-unrolled}). 
We use a target frequency of $200$\,MHz, approximately the \(F_{\max}\) of the full-SoC with \std LLC.
For each design, we first evaluate a single IDF instance, unrolled and pipelined, and then the IDF wrapper used in the LLC, which instantiates multiple IDFs plus the output-slicing logic needed to produce the cache-index bits.
To match the additional tag-access latency, we pipeline each IDF and its wrapper with the same number of stages as its LLC implementation. 
Because the default \sasscache{} configuration does not specify the IDF latency, we conservatively use three stages, following \scattercache{}, whose QARMA-based IDF has a similar structure).

\def\fplo{65.4}\def\fphi{100}             %
\def\fpcmin{55}\def\fpcmax{100}          %
\newcommand{\fpct}[1]{%
	\pgfmathtruncatemacro{\fptint}{round(max(\fpcmin,min(\fpcmax,\fpcmin+(#1-(\fplo))/((\fphi)-(\fplo))*(\fpcmax-\fpcmin))))}%
	\textcolor{blue!\fptint}{(#1\%)}}
\begin{table}[t]
\centering
\caption{Critical Path Analysis of SoC variants.}
\label{tab:hardware-frequency-runs}
\footnotesize
\setlength{\tabcolsep}{2.5pt}
\renewcommand{\arraystretch}{1.08}
\begin{adjustbox}{max width=\columnwidth}
\begin{tabular}{@{}llr@{\,}rlr@{}}
\toprule
& SoC Variant (LLC) & \multicolumn{2}{c}{\(F_{\max}\) (MHz)} & SoC Critical Path & Logic/Route Delay (ns) \\
\midrule
\multirow{5}{*}{\rotatebox{90}{Initial}}
& \std & 206.0 & \fpct{100.0} & MSHR $\rightarrow$ Data-RAM & 1.190/3.702 \\
& \ceasers~\cite{qureshi2019new} & 169.3 & \fpct{82.2} & FSM $\rightarrow$ Metadata-RAM & 1.669/3.972 \\
& \scattercache~\cite{werner2019scattercache} & 203.8 & \fpct{98.9} & MSHR $\rightarrow$ Data-RAM & 1.236/3.677 \\
& \sasscache~\cite{giner2023scatter} & 194.6 & \fpct{94.5} & MSHR $\rightarrow$ Data-RAM & 1.066/4.198 \\
& \mirage~\cite{saileshwar2021mirage} & 134.7 & \fpct{65.4} & Metadata-RAM $\circlearrowleft$ & 2.352/4.523 \\

\multirow{3}{*}{\rotatebox{90}{Redesign}}
&\rule{0pt}{11pt}\ceasersf & 192.5 & \fpct{93.5} & FSM $\rightarrow$ Metadata-RAM & 1.286/3.624 \\
& \sasscachef & 208.6 & \fpct{101.3} & Data-RAM $\rightarrow$ Read Unit & 2.559/1.947 \\
& \miragef & 185.2 & \fpct{89.9} & Hit\_reg $\rightarrow$ Metadata-RAM & 1.385/3.901 \\
\bottomrule
\end{tabular}
\end{adjustbox}
\end{table}

As shown in \cref{tab:idf-pipeline-unrolled}, pipelining substantially shortens the critical paths of both the standalone IDFs  and their corresponding wrappers.
The wrappers generally remain slower because they include multiple IDF instances and the slicing logic, except for \ceasers, whose wrapper is slightly faster owing to a more favorable post-route implementation, where a $0.113$\,ns reduction in routing delay outweighs the additional $0.049$\,ns logic delay.
This inversion reflects single-point physical implementation and WNS extrapolation, rather than a higher intrinsic wrapper \(F_{\max}\).
More importantly, the results show pipelining is a prerequisite rather than an optimization.
In their unrolled form, three of the four wrappers fail to meet the $200$\,MHz reference target, with \sasscache wrapper reaching only $93.2$\,MHz---less than half the target frequency.
After pipelining, every wrappers exceeds $200$\,MHz, with the lowest being \sasscache at $211.6$\,MHz.

\begin{observation}
	\textbf{\textcolor{obsblue}{Observation 1.}}
	When evaluated in isolation, every pipelined IDF wrapper exceeds the 200\,MHz reference target suggesting IDF alone is not the limiting factor for SoC frequency.
\end{observation}

\subsection{Full-SoC Frequency Degradation}
\label{sec:freq-loss}

The isolated evaluation suggests that, once pipelined, the IDF should not limit the operating frequency of the processor.
We now test whether this conclusion holds after integrating the randomized LLCs into a complete SoC.
Unless otherwise stated, all reported \(F_{\max}\), IPC, and FPGA resource results refer to complete NaxRiscv SoC variants identified by their integrated LLC implementation names.
\Cref{tab:hardware-frequency-runs} (Initial SoC variants) reports the highest estimated post-route clock frequencies\footnote{Post-route timing reflects the achievable operating frequency after the complete FPGA implementation flow.} of the \std SoC variant and the four randomized variants.
Since all variants differ only in their LLC implementation, any change in \(F_{\max}\) is attributable to the integrated randomized cache.

The full-SoC results reveal substantial frequency degradation for three of the four variants.
The \ceasers{}, \sasscache{}, and \mirage{} variants achieve only $82.2\%$, $94.5\%$, and $65.4\%$ of the \std{} \(F_{\max}\), respectively, corresponding to a maximum frequency loss of $34.6\%$. 
In contrast, \scattercache{} retains $98.9\%$ of the baseline frequency and preserves the same critical path as \std{}. 
These results contradict the isolated IDF evaluation. 
Although the pipelined \sasscache{} IDF exceeds the $200$\,MHz reference target in isolation, the integrated SoC still incurs a measurable frequency loss. 
Conversely, \ceasers{} employs the fastest pipelined IDF wrapper ($413.2$\,MHz) yet exhibits the second-largest reduction in \(F_{\max}\). 
The dominant timing bottlenecks therefore arise from the integrated LLC rather than the IDF alone. 
To the best of our knowledge, this work provides the first systematic characterization of this full-SoC timing cost.

\begin{observation}
	\textbf{\textcolor{obsblue}{Observation 2.}}
	Pipelining eliminates the IDF as the dominant frequency bottleneck in isolation, yet full-SoC integration still incurs frequency losses of up to $34.6\%$.
\end{observation}

\begin{figure}[t]
	\centering
	\includegraphics[scale=0.7]{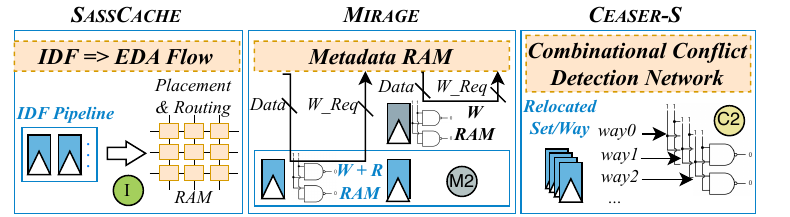}
  \vspace{-2mm}
	\caption{Critical-path characterization of the frequency degradation induced by the \ceasers, \sasscache, and \mirage LLC implementations.}
	\label{fig:freq_cha}
\end{figure}

\subsection{Diagnosing the Critical Paths}
\label{sec:freq-diagnosis}
We now trace the critical paths reported in \cref{tab:hardware-frequency-runs} back to the hardware mechanisms responsible for them.
The resulting critical paths are illustrated in \cref{fig:freq_cha}.

\textbf{\ceasers{}.}
Following the original design, we implement relocation in a non-blocking manner, allowing normal LLC requests to execute concurrently with relocation (cf.~\cref{sec:ceaser-impl}).
To support this, the hazard controller records the set and way of cache lines undergoing relocation and compares them against requests arriving from multiple sources such as the request in input buffer or retry buffer.
As shown in \cref{fig:freq_cha}, these comparisons form a large combinational network within the hazard controller.
In the 16-way LLC configuration, the relocation set and way signals must be broadcast to a large number of comparators, resulting in a high fan-out (125) and increased combinational delay.
Consequently, the \ceasers SoC variant achieves an \(F_{\max}\) of only 169.3~MHz, corresponding to 82.2\% of that of the \std SoC variant.

\textbf{\sasscache{}.}
Although its pipelined IDF exceeds the target frequency in isolation (cf.~\cref{sec:IDF_Iso}),  integrating it into the LLC still reduces the SoC frequency through placement and routing rather than IDF logic. 
The critical path originates at the MSHRs, whose write-side control signals depend on the IDF-derived randomized index in the final controller stage (cf. \cref{fig:cache-structure}). 
Although several pipeline stages separate these components logically, the dense IDF logic biases placement toward the MSHRs, increasing their distance from the data RAMs and hence the routing delay on the MSHR-to-data-RAM path (\cref{fig:freq_cha}).
Consequently, the \sasscache{} SoC reaches only $194.6$\,MHz ($94.5\%$ of \std). 
The \scattercache{} SoC does not exhibit this effect because its simpler five-round QARMA IDF is more evenly distributed across three pipeline stages.

\textbf{\mirage{}.}
To support indirect eviction, \mirage introduces additional metadata ways, the associated metadata structures, and the extra metadata accesses (cf.~\cref{sec:mirage-impl}).
The additional ways widen the comparison and selection networks used for hit detection and metadata updates.
As \cref{fig:freq_cha} shows, both critical paths begin at the metadata RAM, traverse the combinational comparison and update logic, and drive the write data or write request back to the metadata RAM within the same pipeline stage.
This forms long RAM-to-logic-to-RAM read--modify--write paths.
These additional accesses and the enlarged metadata logic form \mirage{}'s dominant frequency bottleneck, achieving only $134.7$\,MHz, with logic accounting for $34.2\%$ of the critical-path delay and routing contributing the remaining $65.8\%$.

Across all three degraded variants, routing rather than logic accounts for the majority of the critical-path delay (65.8--79.7\%, \cref{tab:hardware-frequency-runs}).
The frequency cost therefore arises primarily from fan-out, placement, and wiring---physical-design effects that are not exposed by isolated component analysis.

\begin{observation}
	\textbf{\textcolor{obsblue}{Observation 3.}}
	The frequency bottlenecks arise from three distinct mechanisms: concurrent-relocation conflict handling in \ceasers, placement-and-routing pressure from IDF integration in \sasscache, and additional accesses to enlarged metadata RAMs in \mirage.
	No single mitigation applies to all designs.
\end{observation}

\subsection{Restoring Frequency by Redesign}
\label{sec:freq-restore}

Having identified the sources of the frequency degradation, we now redesign the LLC logic along the critical paths of the three affected SoC variants.
We propose the frequency-optimized LLC implementations \ceasersf, \sasscachef, and \miragef and integrate each into an otherwise unchanged SoC.\footnote{We do not further optimize \scattercache{} as its frequency is already close to the standard SoC variant.}
The resulting timing improvements are summarized in \cref{tab:hardware-frequency-runs} (Redesign SoC variants).

For \ceasers, we propose \ceasersf, which disables concurrent request processing during relocation.
Each relocation operation temporarily stalls incoming LLC requests, eliminating the need to compare the relocation state against requests from multiple sources. 
This simplifies the comparison network (\cref{fig:freq_cha}) and reduces its fan-out from $125$ to $46$.
Consequently, the \ceasersf SoC variant achieves a maximum operating frequency of $192.5$\,MHz, a $13.7$\% improvement over the original \ceasers variant.
Although the critical path remains the same, its overall delay is substantially reduced by the lower fan-out.

For \sasscache, we propose \sasscachef, which increases the number of IDF pipeline depth from three to four stages.
The additional stage provides the placer with greater flexibility in the IDF region, allowing the MSHRs to be placed closer to the data RAM and thereby reducing routing delay.
After this optimization, the \sasscachef SoC variant reaches $208.6$~MHz, exceeding the \std SoC frequency.
This improvement may be attributed to the dedicated metadata-RAM access stage in the randomized LLC implementation, described in \cref{sec:common-impl} (\circnum{I}{idf}), which can improve placement and routing.
Additionally, the critical path no longer spans the MSHRs and data RAM, instead shifting entirely into the backend pipeline.

Finally, for \mirage, we redesign the parallel pipeline associated with \circnum{M2}{mirage} in \cref{fig:cache-structure}.
Rather than overlapping the additional metadata access with the controller pipeline, \miragef introduces two pipeline stages to buffer the metadata-RAM response.
These stages pipeline the two long combinational paths indicated by the black arrows in \cref{fig:freq_cha}.
After optimization, the \miragef SoC variant achieves 185.2~MHz, corresponding to $89.9$\% of the frequency of the \std SoC variant.
The critical path shifts to the registers that retain cache-hit information for scheduling in the subsequent cycle, located in Stage~2 of the control pipeline (cf.~\cref{fig:cache-structure}).
The remaining delay primarily arises from the architectural characteristics of the \mirage LLC, including its $75\%$ increase in the number of ways and the resulting increase in combinational-logic complexity.

\begin{observation}
	\textbf{\textcolor{obsblue}{Observation 4.}}
	Most of the lost frequency is recoverable (89.9--101.3\% of the \std \(F_{\max}\)), but only through design-specific trade-offs: reduced concurrency in \ceasersf{}, increased IDF latency in \sasscachef{}, and additional pipeline stages in \miragef{}.
\end{observation}

\section{Evaluation}
\label{sec:evaluation}

We evaluate the complete SoC variants by measuring application performance together with FPGA implementation and ASIC synthesis results, using instructions per cycle (IPC) as the primary performance metric. Unless otherwise stated, all variants differ only in their LLC implementation, and the experimental setup is summarized below.

\begin{table}[H]
	\centering
	\footnotesize
	\textbf{Experimental Setup}\\[2pt]
	\footnotesize
	\setlength{\tabcolsep}{8pt}
	\renewcommand{\arraystretch}{1.25}
	\begin{tabular}{@{}
			>{\raggedright\arraybackslash}m{0.14\columnwidth}
			>{\raggedright\arraybackslash}p{0.80\columnwidth}
			@{}}
		\toprule
		SoC
		& NaxRiscv~\cite{spinalhdl2026naxriscv}, one or two out-of-order cores at 125~MHz;
		32~KiB L1I/L1D; 2~MB 16-way L2 as the LLC; 2~GB DDR4 at 500~MHz.
		\\
		Software
		& Benchmarks are compiled with the RISC-V GCC toolchain~\cite{jelinek2024gcc}, using \texttt{-fopenmp} for the multicore runs.
		\\
		Benchmarks
		& \emph{Single core:} SPEC CPU 2017 rate~\cite{spec2017spec}:
		all 23 int and FP programs.
		GAP~\cite{beamer2015gap}: \bench{TC}, \bench{BFS}, \bench{CC},
		\bench{SSSP}, \bench{BC}, \bench{PR}.
		\newline
		\emph{Multicore:} four most memory-intensive programs of each suite~\cite{ranjan2019spec}---SPEC CPU 2017 speed:
		\bench{for}, \bench{rom}, \bench{bwa}, \bench{lbm};
		GAP: \bench{BC}, \bench{PR}, \bench{CC}, \bench{SSSP}.
		\\
		FPGA
		& AMD Xilinx VCU118 (Virtex UltraScale+)~\cite{amd2023vcu118} with Vivado 2020.2.
		Resource utilization is taken from post-implementation reports, and dynamic power from Vivado's vector-based analysis flow~\cite{amd2026vivado}\tablefootnote{As per AMD documentation, power analysis at post-route provides the most accurate power estimation~\cite{amd2026vivado}.}.
		\\
		ASIC
		& Synopsys Design Compiler with the Nangate 45~nm standard-cell library~\cite{nangate2008open} and OpenRAM SRAMs at the same technology node~\cite{guthaus2016openram}.
		\\
		\bottomrule
	\end{tabular}
\end{table}

\begin{figure}[t]
    \centering
	    \includegraphics[width=0.98\linewidth]{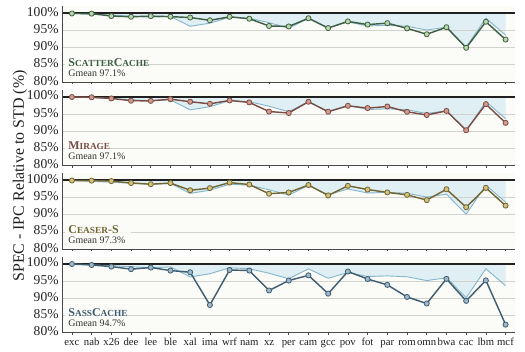}
	    \vspace{-1mm}
	    \caption{\(\mathrm{IPC}\) of the four initial randomized-LLC SoC variants normalized to \std for SPECrate CPU 2017. Labels are abbreviated to the first three letters.}
	    \vspace{-3mm}
    \label{fig:corr-ipc}
\end{figure}

\subsection{Performance}
\label{sec:eval-perf}

\paragraph{Baseline Characterization}
\Cref{fig:corr-ipc} reports the IPC of each initial SoC variant with a randomized LLC, normalized to the \std SoC variant.
Programs are ordered from left to right by increasing accesses per kilo-instruction (\(\mathrm{APKI}\)), illustrating how LLC access intensity influences performance.\footnote{Although our LLC corresponds to the second cache level, its average access rate (40 APKI) is close to the average L2 miss rate (36 MPKI) reported for SPEC CPU2017 on an Intel Core i7-8700K~\cite{ranjan2019spec}, suggesting that our LLC experiences a representative volume of requests.}
On average, the \scattercache and \mirage SoC variants reduce IPC by $2.9\%$ on SPEC and $6.3\%$ on GAP because both introduce the same additional control-pipeline latency. 
In contrast, \ceasers{} incurs one fewer pipeline stage, reducing the average IPC loss to $2.7\%$ on SPEC and $6.1\%$ on GAP.
The \sasscache{} SoC variant exhibits the largest IPC reduction---$5.3\%$ on SPEC and $11.3\%$ on GAP---because it combines the additional control-pipeline latency with a higher LLC miss rate resulting from its reduced effective cache capacity.
Interestingly, the multicore results follow the same trend as the single-core, except for \sasscache{}, whose reduced effective cache capacity increases the IPC degradation by up to $9\%$. 
While these results quantify the overall performance impact, they do not explain its underlying cause. 
We therefore next examine whether randomization-induced LLC misses account for the observed slowdown.

\paragraph{Miss-Rate Analysis}
Following prior work~\cite{qureshi2019new,werner2019scattercache,saileshwar2021mirage,giner2023scatter}, we consider two potential sources of IPC degradation: randomization-induced LLC misses and the additional control-pipeline latency incurred on every LLC access.
For each benchmark, \(\Delta\mathrm{MPKI}\) measures the \%age increase in LLC misses per kilo-instruction over \std, whereas \(\mathrm{APKI}\) measures how frequently the additional latency is incurred.
\Cref{tab:delta-mpki-ipc-correlation} directly contrasts the Pearson correlations of normalized IPC with these two metrics.
For SPEC, the \(\Delta\mathrm{MPKI}\)--\(\mathrm{IPC}\) correlations are weak and slightly positive ($0.086$--$0.325$), whereas the \(\mathrm{APKI}\)--\(\mathrm{IPC}\) correlations are consistently negative and stronger ($-0.803$ to $-0.717$).
Thus, additional misses alone do not explain the SPEC degradation, which instead tracks LLC access intensity.
For GAP, both \(\Delta\mathrm{MPKI}\) ($-0.860$ to $-0.651$) and \(\mathrm{APKI}\) ($-0.927$ to $-0.882$) correlate strongly with normalized \(\mathrm{IPC}\).
Correlation, however, characterizes how two metrics vary together across benchmarks, not the magnitude of their contribution. To provide this context, we also examine the average increase in LLC misses. 
Relative to the suite-mean \std{}  \(\mathrm{MPKI}\) , the suite-mean \(\Delta\mathrm{MPKI}\) of \scattercache{}, \mirage{}, and \ceasers{} is only $3.1$--$8.7\%$ for SPEC and $6.5$--$12.3\%$ for GAP. 
Only \sasscache{} exhibits a substantial \(\mathrm{MPKI}\) increase because its reduced effective cache capacity increases LLC misses.
These statistics alone cannot identify the dominant source of IPC degradation. 
We therefore isolate the per-access latency cost using latency-matched controls.

\begin{table}[t]
	\centering
	\caption{MPKI statistics and Correlations with normalized \(\mathrm{IPC}\).}
	\label{tab:delta-mpki-ipc-correlation}
	\scriptsize
	\setlength{\tabcolsep}{5pt}
	\renewcommand{\arraystretch}{0.95}
	\resizebox{\columnwidth}{!}{%
		\begin{tabular}{@{}l *{8}{r}@{}}
			\toprule
			& \multicolumn{4}{c}{SPEC} & \multicolumn{4}{c}{GAP} \\
			\cmidrule(lr){2-5}\cmidrule(l){6-9}
			& \cache{Sct} & \cache{Mir} & \cache{Ces} & \cache{Sas}
			& \cache{Sct} & \cache{Mir} & \cache{Ces} & \cache{Sas} \\
			\midrule
			\(\Delta\mathrm{MPKI}\)
			& +3.1\% & +3.1\% & +8.7\% & +53.4\%
			& +6.5\% & +6.5\% & +12.3\% & +42.3\% \\
			\midrule
			\(r_{\Delta\mathrm{MPKI},\mathrm{IPC}}\)
			& 0.086 & 0.087 & 0.325 & 0.228
			& -0.663 & -0.651 & -0.848 & -0.860 \\
			\(r_{\mathrm{APKI},\mathrm{IPC}}\)
			& -0.803 & -0.774 & -0.771 & -0.717
			& -0.890 & -0.889 & -0.927 & -0.882 \\
			\bottomrule
		\end{tabular}%
	}
	
	\vspace{2pt}
	
	\begin{minipage}{\columnwidth}
		\scriptsize
		\emph{Note:} \cache{Sct}: \scattercache; \cache{Mir}: \mirage; \cache{Ces}: \ceasers; \cache{Sas}: \sasscache.
	\end{minipage}
	\vspace{-2mm}
\end{table}
\begin{figure}[t]
	\centering
	\includegraphics[scale=1]{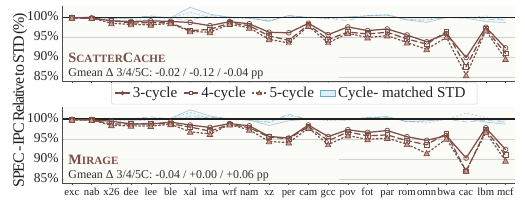}
  \vspace{-4mm}
	\caption{Latency-matched SPEC validation for \scattercache and \mirage with varying IDF latencies. The blue-shaded part denotes the \(\mathrm{IPC}\) degradation of latency-matched \std variant.}
	\vspace{-3mm}
	\label{fig:latency-validation}
\end{figure}

\paragraph{Latency-Matched Validation}

To isolate the per-access latency, we construct \stda by adding the default three-cycle LLC latency of \scattercache and \mirage to \std using pipeline registers but no IDF.
The remaining IPC difference captures non-latency effects, including randomization-induced misses.
Across both suites, \stda closely matches \scattercache, \mirage, and the one-cycle-shorter \ceasers.
Their geometric-mean \(\mathrm{IPC}\) differs by at most $0.2$ percentage points on SPEC and $1.3$ on GAP.
Compared with the total IPC degradations of roughly $3\%$ and $6\%$, these residuals show that latency dominates the performance overhead.
\sasscache falls further because its reduced effective capacity increases LLC misses.
We further compare four-, and five-cycle \scattercache and \mirage configurations against latency-matched \std variants.
\Cref{fig:latency-validation} shows that the SPEC residual remains within $0.12$ percentage points for both caches at every evaluated pipeline depth.

\begin{observation}
\textbf{\textcolor{obsblue}{Observation 5.}}
Except for \sasscache{}, IPC degradation tracks LLC access intensity rather than LLC misses. Latency-matched evaluation confirm that additional LLC access latency dominates the performance overhead of randomized caches.
\end{observation}

\subsection{Maximum Execution Speed}
\begin{table}[t]
    \centering
    \caption{IPC, maximum execution speed (MIPS), and MIPS slowdown.}
    \label{tab:ipc-mips-spec-gap}

    \newcommand{\slowcell}[2]{%
        \pgfmathtruncatemacro{\slowtint}{round(5+min(1,(#1)/(#2))*43)}%
        \edef\slowtmp{\noexpand\cellcolor{orange!\slowtint}}\slowtmp #1\%}

    \begin{adjustbox}{max width=\columnwidth}
        \begin{tabular}{@{}l *{6}{r}@{}}
            \toprule
            & \multicolumn{3}{c}{SPEC}
            & \multicolumn{3}{c}{GAP} \\
            \cmidrule(lr){2-4}
            \cmidrule(l){5-7}

            SoC Variant
            & IPC & MIPS & Slowdown
            & IPC & MIPS & Slowdown \\
            \midrule

            \std
            & 0.692 & 135.58 & --
            & 0.348 & 68.19 & -- \\

            \ceasers
            & 0.683 & 109.95 & \slowcell{18.90}{34.85}
            & 0.342 & 55.08 & \slowcell{19.23}{32.43} \\

            \scattercache
            & 0.672 & 130.35 & \slowcell{3.86}{34.85}
            & 0.327 & 63.35 & \slowcell{7.09}{32.43} \\

            \sasscache
            & 0.648 & 119.94 & \slowcell{11.54}{34.85}
            & 0.305 & 56.46 & \slowcell{17.21}{32.43} \\

            \mirage
            & 0.690 & 88.34 & \slowcell{34.85}{34.85}
            & 0.360 & 46.08 & \slowcell{32.43}{32.43} \\

            \midrule

            \ceasersf
            & 0.674 & 123.42 & \slowcell{8.97}{34.85}
            & 0.330 & 60.39 & \slowcell{11.45}{32.43} \\

            \sasscachef
            & 0.638 & 126.38 & \slowcell{6.79}{34.85}
            & 0.295 & 58.45 & \slowcell{14.29}{32.43} \\

            \miragef
            & 0.667 & 117.39 & \slowcell{13.41}{34.85}
            & 0.327 & 57.64 & \slowcell{15.47}{32.43} \\

            \bottomrule
        \end{tabular}
    \end{adjustbox}

    \vspace{2pt}

    \begin{minipage}{\columnwidth}
        \scriptsize
        \emph{Note:} IPC is calculated as Gmean of all benchmarks running under each design's \(F_{\max}\). MIPS is IPC multiplied by \(F_{\max}\).
    \end{minipage}
    \vspace{-5mm}
\end{table}

IPC alone does not capture the overall performance impact of randomized LLCs because it ignores the reduction in processor frequency. 
We therefore evaluate the maximum execution speed under each design's maximum reachable frequency, computed as \(\mathrm{IPC}\) $\times$ \(F_{\max}\), and report it in millions of instructions per second (MIPS). 
As shown in \cref{tab:ipc-mips-spec-gap}, incorporating the frequency reveals a much larger performance loss.
In general, because the memory frequency is fixed, a lower execution frequency translates to a reduction in the number of cycles a cache miss costs, which somewhat compensates for the additional latency that randomized caches induce.
However, although the IPC loss is reduced to at most $6\%$, the overall execution speed still drops by up to almost $35\%$, once the maximal execution fequency is taken into account.  
The frequency-optimized implementations recover most of this loss.
In particular, \miragef{} improves the slowdown from $34.85\%$ to $13.41\%$ on SPEC and from $32.43\%$ to $15.47\%$ on GAP, despite incurring a small additional IPC penalty.
These results show that evaluating IPC in isolation substantially underestimates the true performance cost of randomized caches.

\subsection{Resource Utilization and Power}

\Cref{tab:hardware-area-resources} reports post-implementation FPGA resource utilization and dynamic power together with post-synthesis ASIC area, all normalized to \std. Among the initial designs, \scattercache{} and \sasscache{} incur similar overheads because they use similar IDFs (about $10\%$ LUTs, $6\%$ FFs, and $1.4\%$ BRAM). \ceasers{} exhibits the largest LUT overhead ($\sim 35\%$) due to relocation and conflict-handling logic, although mapping metadata to UltraRAM reduces BRAM utilization by $3.5\%$. \mirage{} increases LUT, FF, and BRAM utilization by about $15\%$, $11\%$, and $11\%$, respectively, owing to its additional metadata ways.

The redesigns preserve the randomized-cache algorithms while shortening their critical paths. Simplifying relocation lowers the LUT overhead of \ceasersf from $35.2\%$ to $18.9\%$ with essentially unchanged FF utilization. 
The additional IDF stage in \sasscachef adds only about one percentage point of FF overhead and leaves LUT and BRAM utilization nearly unchanged. 
In contrast, the deeper pipeline in \miragef raises FF overhead from $11.4\%$ to $29.6\%$, while its LUT and BRAM overheads remain close to those of \mirage.

Unlike FPGA LUT utilization, ASIC area is dominated by the LLC SRAM macros rather than synthesized control logic. 
Consequently, the additional hazard logic, FSMs, and IDFs contribute little to the overall silicon area, yielding SoC area overheads of only $1$--$14\%$. 
Instead, the ASIC area closely follows the FPGA memory overhead---BRAM, together with \ceasers{}' UltraRAM---showing that additional cache storage, rather than control logic, dominates the silicon cost.

Dynamic power overhead ranges from about $12$--$28\%$ at the SoC level and $30$--$75\%$ within the LLC. 
The IDF dominates the power overhead of \scattercache{} and \sasscache{}, whereas \ceasers{} is additionally affected by UltraRAM accesses and the $160$-bit rekeying LFSR. 
\mirage{} incurs higher LLC power because of its expanded metadata structures. 
The redesigns have little impact on SoC-level power. \sasscachef{} reduces power by shortening the critical path, \ceasersf{} remains nearly unchanged because relocation is infrequent, and \miragef{} slightly lowers SoC power despite higher LLC power by increasing processor stalls.

\begin{table}[t]
	\centering
	\caption{Post-implementation FPGA and post-synthesis ASIC results.}
	\label{tab:hardware-area-resources}
	\footnotesize
	\setlength{\tabcolsep}{2.5pt}
	\renewcommand{\arraystretch}{1.10}
	
	\def\hwlo{-3.5}\def\hwhi{75.1}      %
	\def\hwcmin{5}\def\hwcmax{48}       %
	\makeatletter
	\def\hwsetpct#1\%#2\hwend{\def\hwpct{#1}}
	\makeatother
	\newcommand{\hwcell}[2]{%
		\hwsetpct#2\hwend
		\pgfmathtruncatemacro{\hwtint}{round(max(\hwcmin,min(\hwcmax,\hwcmin+(\hwpct-(\hwlo))/((\hwhi)-(\hwlo))*(\hwcmax-\hwcmin))))}%
		\edef\hwtmp{\noexpand\cellcolor{orange!\hwtint}}\hwtmp
		\begin{tabular}[t]{@{}r@{}}%
			#1\\[-2pt]%
			{\tiny(#2)}%
	\end{tabular}}
	
	\begin{adjustbox}{max width=\columnwidth}
		\begin{tabular}{@{}l *{7}{Y}@{}}
			\toprule
			
			&
			\multicolumn{3}{c}{FPGA Area}
			&
			\multicolumn{2}{c}{Dyn Power}
			&
			\multicolumn{2}{c}{ASIC Area}
			\\
			
			\cmidrule(lr){2-4}
			\cmidrule(lr){5-6}
			\cmidrule(l){7-8}
			
			SoC Variant
			&
			LUT
			&
			{FF \;}
			&
			BRAM
			&
			SoC
			&
			LLC
			&
			SoC
			&
			LLC
			\\
			
			&
			&
			&
			&
			(mW)
			&
			(mW)
			&
			(mm$^2$)
			&
			(mm$^2$)
			\\
			
			\midrule
			
			\std
			&
			\cellcolor{gray!20}40,699
			&
			\cellcolor{gray!20}29,221
			&
			\cellcolor{gray!20}579.5
			&
			\cellcolor{gray!20}654
			&
			\cellcolor{gray!20}238
			&
			\cellcolor{gray!20}57.757
			&
			\cellcolor{gray!20}54.903
			\\
			\ceasers
			&
			\hwcell{55,021}{+35.2\%}
			&
			\hwcell{33,074}{+13.2\%}
			&
			\hwcell{559.0}{-3.5\%}
			&
			\hwcell{738}{+12.8\%}
			&
			\hwcell{326}{+36.6\%}
			&
			\hwcell{59.007}{+2.2\%}
			&
			\hwcell{56.153}{+2.3\%}
			\\
			
			\scattercache
			&
			\hwcell{44,611}{+9.6\%}
			&
			\hwcell{30,972}{+6.0\%}
			&
			\hwcell{587.5}{+1.4\%}
			&
			\hwcell{732}{+11.9\%}
			&
			\hwcell{309}{+29.8\%}
			&
			\hwcell{58.417}{+1.1\%}
			&
			\hwcell{55.564}{+1.2\%}
			\\
			
			\sasscache
			&
			\hwcell{44,708}{+9.9\%}
			&
			\hwcell{30,894}{+5.7\%}
			&
			\hwcell{587.5}{+1.4\%}
			&
			\hwcell{840}{+28.4\%}
			&
			\hwcell{417}{+75.1\%}
			&
			\hwcell{58.417}{+1.1\%}
			&
			\hwcell{55.564}{+1.2\%}
			\\
			
			\mirage
			&
			\hwcell{46,627}{+14.6\%}
			&
			\hwcell{32,542}{+11.4\%}
			&
			\hwcell{642.5}{+10.9\%}
			&
			\hwcell{798}{+21.9\%}
			&
			\hwcell{361}{+51.4\%}
			&
			\hwcell{65.854}{+14.0\%}
			&
			\hwcell{63.000}{+14.7\%}
			\\
			\midrule
			\ceasersf
			&
			\hwcell{48,408}{+18.9\%}
			&
			\hwcell{33,100}{+13.3\%}
			&
			\hwcell{559.0}{-3.5\%}
			&
			\hwcell{751}{+14.8\%}
			&
			\hwcell{327}{+37.4\%}
			&
			\hwcell{58.982}{+2.1\%}
			&
			\hwcell{56.128}{+2.2\%}
			\\
			
			\sasscachef
			&
			\hwcell{44,586}{+9.6\%}
			&
			\hwcell{31,163}{+6.6\%}
			&
			\hwcell{587.5}{+1.4\%}
			&
			\hwcell{719}{+9.8\%}
			&
			\hwcell{310}{+30.1\%}
			&
			\hwcell{58.419}{+1.1\%}
			&
			\hwcell{55.565}{+1.2\%}
			\\
			
			\miragef
			&
			\hwcell{47,251}{+16.1\%}
			&
			\hwcell{37,883}{+29.6\%}
			&
			\hwcell{642.5}{+10.9\%}
			&
			\hwcell{795}{+21.6\%}
			&
			\hwcell{381}{+59.9\%}
			&
			\hwcell{65.889}{+14.1\%}
			&
			\hwcell{63.035}{+14.8\%}
			\\
			\bottomrule
		\end{tabular}
	\end{adjustbox}
	
	\vspace{2pt}
	
	\begin{minipage}{\columnwidth}
		\scriptsize
		\emph{Note:} FPGA results are for the complete SoC; Power measurement excludes I/O
	\end{minipage}
	    \vspace{-3mm}
\end{table}

\section{Conclusion and Future Works}
\label{sec:discussion}

Randomized caches are among the most promising defenses against cache side-channel attacks, yet their practical deployment remains largely unexplored.
In this work, we presented the first end-to-end hardware study of state-of-the-art randomized caches on a high-performance out-of-order RISC-V SoC.
We identified the implementation challenges of integrating cache randomization into modern non-blocking LLCs, developed corresponding hardware solutions, and evaluated four representative designs.
Our results revealed a previously overlooked processor-frequency cost of randomized caches, traced it to its underlying critical paths, and showed that most of this loss can be recovered through targeted microarchitectural redesigns. 
Together, these findings establish a practical foundation for implementing randomized caches and provide the hardware insights needed to evaluate future randomized-cache designs under realistic implementation constraints.

The insights gained from this study suggest several directions for future work.
Although our study targets an advanced open-source out-of-order RISC-V SoC, commercial processors feature larger cache hierarchies, deeper pipelines, and more complex memory systems. Extending this analysis to such platforms is therefore an important next step. Likewise, many randomized-cache proposals beyond the four studied here remain unevaluated in hardware, and implementing them would further broaden our understanding of the randomized-cache design space. Finally, completing a full ASIC implementation flow, including physical design, would provide additional insight into optimization opportunities and bottlenecks.

\section*{Acknowledgments}
This work has been supported by 
  the Advanced Research Projects Agency for Health (ARPA-H) under Agreement No.\ 140D042590046; 
  the Alfred P. Sloan Research Fellowship;
  an ARC Discovery Project number DP210102670; 
  the Deutsche Forschungsgemeinschaft (DFG, German Research Foundation) under Germany's Excellence Strategy - EXC 2092 CASA - 390781972 
  and through project number 560392681; 
  a grant from Coefficient Giving; and gifts from Qualcomm and Zama. 
\bibliographystyle{IEEEtranSN}
\bibliography{Bib}

\end{document}